# Preventing Incomplete/Hidden Requirements: Reflections on Survey Data from Austria and Brazil

Marcos Kalinowski[1], Michael Felderer[2], Tayana Conte[3], Rodrigo Spínola[4],
Rafael Prikladnicki[5], Dietmar Winkler[6], Daniel Méndez Fernández[7], Stefan Wagner[8]

[1] Universidade Federal Fluminense, Computing Institute, Av. Milton Tavares de Souza s/n, Campus Praia Vermelha, 24210-346 Niterói, Brazil.
[2] University of Innsbruck, Institute of Computer Science, Technikerstr. 21a, A-6020 Innsbruck, Austria.
[3] Universidade Federal do Amazonas, Computing Institute, Av. Rodrigo Otávio 6200, Campus Universitário Senador Arthur Virgílio Filho, 69077-000, Manaus, Brazil
[4] Universidade Salvador, Systems and Computing Graduate Programm, Alameda das Espatódias 912, 41.820-460, Salvador, Brazil.
[5] Pontifícia Universidade Católica do Rio Grande do Sul, Computer Science Graduate Programm, Av. Ipiranga 6681, 90619-900 Porto Alegre, Brazil.
[6] Vienna University of Technology, Institute of Software Technology & Interactive Systems, Favoritenstr. 9/188, A-1040 Vienna, Austria.
[7] Technische Universität München, Institut für Informatik, Boltzmannstr. 3 85748 Garching, Germany
[8] University of Stuttgart, Institut für Softwaretechnologie, Universitätsstraße 38 D-70569 Stuttgart, Germany

[1]kalinowski@ic.uff.br, [2]michael.felderer@uibk.ac.at, [3]tayana@icomp.ufam.edu.br, [4]rodrigo.spinola@pro.unifacs.br, [5]rafael.prikladnicki@pucrs.br, [6]dietmar.winkler@tuwien.ac.at, [7]daniel.mendez @tum.de, [8]stefan.wagner@informatik.uni-stuttgart.de

**Abstract.** [Context] Many software projects fail due to problems in requirements engineering (RE). [Goal] The goal of this paper is analyzing a specific and relevant RE problem in detail: incomplete/hidden requirements. [Method] We replicated a global family of RE surveys with representatives of software organizations in Austria and Brazil. We used the data to (a) characterize the criticality of the selected RE problem, and to (b) analyze the reported main causes and mitigation actions. Based on the analysis, we discuss how to prevent the problem. [Results] The survey includes 14 different organizations in Austria and 74 in Brazil, including small, medium and large sized companies, conducting both, plan-driven and agile development processes. Respondents from both countries cited the incomplete/hidden requirements problem as one of the most critical RE problems. We identified and graphically represented the main causes and documented solution options to address these causes. Further, we compiled a list of reported mitigation actions. [Conclusions] From a practical point of view, this paper provides further insights into common causes of incomplete/hidden requirements and on how to prevent this problem.

# 1  Introduction

The importance of high-quality requirements engineering (RE) has been widely accepted and well documented. RE constitutes a holistic key to successful development projects [1]. However, industry is still struggling to apply high-quality RE practices [2] and getting a further understanding on common RE problems and their causes is of great interest to both, industry and academy.

Many researchers have addressed identifying and analyzing RE problems faced by industry [3][4]. More recently, a project called NaPiRE (**Na**ming the **P**ain **i**n **R**equirements **E**ngineering) comprises the design of a family of surveys on RE practice and problems, and it is conducted in joint collaboration with various researchers from different countries [5]. The main goal of this project is to provide an empirical foundation on the state of the practice in RE to allow steering future research in a problem-driven manner. The NaPiRE survey includes several countries around the globe[1].

From the perspective of practitioners, information on RE problems could be particularly useful to discuss how to prevent the occurrence of such problems in their projects. An efficient means for preventing RE problems is the causal analysis [6], which involves identifying causes of problems to address them through concrete actions to prevent them in future projects. Kalinowski *et al.* [7] provide a comprehensive industrial experience report on conducting causal analysis on RE problems. One of the main difficulties reported during causal analysis sessions concerns the absence of a starting point for identifying potential causes [6], as there is no general documented and empirically grounded knowledge on common causes of critical RE problems usable as a starting point.

Data collected in the NaPiRE survey include information on critical RE problems and their causes. An initial effort to organize knowledge on common causes of critical RE problems has been recently undertaken based on NaPiRE data from the Brazilian replication [8]. In this paper, we extend this research by further analyzing a specific and critical selected RE problem: incomplete/hidden requirements, based on data from the NaPiRE replications conducted in Austria and Brazil. We use the data to (a) characterize the criticality of the selected RE problem, and to (b) analyze the main causes reported for the problem. Based on this industrial feedback, we discuss actions for preventing the problem. As a result of the replications, we received complete answers from 14 different organizations in Austria and 74 in Brazil, including small, medium and very large sized companies, conducting both, plan-driven and agile development. Respondents from both countries cited the selected problem as one of the most critical RE problems. We graphically represent the causes cited by the organizations and discuss solution options for addressing the most common reported causes.

The remainder of this paper is organized as follows. Section 2 describes the background on surveys on RE problems and on the NaPiRE project. Section 3 describes the NaPiRE survey replication in Austria and in Brazil. Section 4 presents the survey results on the criticality of RE problems in both countries. Section 5 contains the analysis of the selected problem including its main reported causes and the discussion on solution options for addressing them. Finally, Section 6 presents the concluding remarks and future work.

---

[1] NaPiRE: http://www.re-survey.org

## 2 Background

As background for this paper, we describe related work on surveys on RE problems and the required information on the NaPiRE project.

### 2.1 RE Surveys

A well-known survey on causes for project failure is the Chaos Report of the Standish Group on cross-company root causes for project failures. While most of these causes are related to RE, the survey has serious design flaws and the validity of its results is questionable [9]. Additionally, it exclusively investigated failed projects and general causes at the level of overall software projects. Thus, unfortunately it does not directly support the investigation of RE problems in industry.

Some surveys have been focusing specifically on RE problems in industry. These surveys include the one conducted by Hall *et al.* [3] in twelve software organizations. Their findings, among others, suggest that most RE problems are organizational rather than technical. Some country-specific RE problem investigations include the surveys conducted by Solemon *et al.* [10] and Liu *et al.* [11], with Malaysian and Chinese organizations, respectively. Khankaew and Riddle [12], report on a survey with focus on more recently conducted semi-structured interviews with organizations from Thailand. These investigations provide valuable insights into industrial environments. However, as each of them focuses on specific aspects in RE, their results are isolated and not generalizable. To address this issue, the NaPiRE project was launched in a joint collaboration with researchers from different countries [5].

### 2.2 The NaPiRE Project

The NaPiRE project resulted in the design of a global family of surveys to overcome the problem of isolated investigations in RE that are not representative [5]. Thus, a long-term goal of the project is to establish an empirically sound basis for understanding trends and problems in RE [13]. Currently several surveys are going to be replicated in several countries around the globe.

The design of the survey is aligned to a well-thought theory and its instruments have been extensively reviewed by several researchers [5][13]. In summary, the NaPiRE survey contains 35 questions with focus on the following type of data from the responding organizations: (a) general information, (b) RE status quo, (c) RE improvement status quo, (d) RE problems faced in practice, and (e) RE problem manifestation (e.g., causes and impact). Further information on the project is available online[1], including the target countries for survey replication and a sample of the questionnaire. Up to now, initial results from Germany have already been published [5][13]. Currently, these initial results will now be updated by more recent trials in Germany and in other countries, such as Austria and Brazil.

## 3 Replicating the NaPiRE Survey in Austria and Brazil

This section describes the collected data in context of this paper based on the Austrian and Brazilian replication. Note that both replications apply the common design of the NaPiRE survey, including all relevant instruments (see [5] for details). Therefore, in this section we focus on the details on planning and execution aspects in both countries, i.e. in Austria and Brazil. To enable proper interpretations of the results, we include a description of the characterization of the responding organizations of both countries in this section.

### 3.1 Survey Replication in Austria

The Austrian NaPiRE survey replication was planned in two meetings with the general NaPiRE organizers from Germany. During these meetings, the online environment (EFS survey tool[2]) was introduced and some guidelines for conducting the survey were presented. For the survey in Austria, the questionnaire, applied in Germany, was duplicated and hosted on the same online environment.

As the goal of the survey was to gain high quality feedback on topics related to RE, the invitation to participate in the survey was sent – in coordination with the general NaPiRE organizers – to selected experts in requirements and software quality engineering of representative organizations in Austria. The organizations covered development of embedded as well as information systems in different domains.

Invitation letters, including a link to the online survey and a password, were sent to the list of experts via e-mail in June 2014 and July 2014. In total, 22 of 25 invited experts logged into the online survey and provided answers between June and September 2014. Out of these, we received 14 completed surveys; 8 experts dropped the survey before completion. The median duration for completing the survey was about 30 minutes.

### 3.2 Survey Replication in Brazil

The planning of the survey replication in Brazil also involved two meetings with the NaPiRE general organizers[1]. Again, during these meetings, the online environment (EFS survey tool[2]) was presented and some general guidelines for conducting the survey were provided. For this replication we decided to translate all instruments to Portuguese, the participants' native language.

Given the geographic dimensions of Brazil, to reach organizations from different regions and to collect representative data, the first author assembled a team of industry-focused researchers spread across the country. The strategy consisted of having researchers from the four main industry intensive regions of the country involved. The resulting NaPiRE Brazil team[1] comprises a researcher from the South of the country, one from the Southeast, one from the North and one from the Northeast. Additionally, we contacted Softex[3], the association responsible for the most widely adopted software process improvement reference model in Brazil, the MPS-SW[4] [14], with over 600 assessments in Brazil. They promptly trusted us contacts of 254 organizations with currently valid MPS-SW assessments so that they could be invited to take part in the survey. Including a set of 80 additional relevant industry contacts from the authors (20 contacts per author on average), we created a list with contacts of representatives from 334 software organizations. We believe this set to be representative for the Brazilian software industry. Given the size of this industry (thousands of software organizations [15]), an extensive survey to reach all of them would be almost impossible. We then configured the environment and sent the invitations with a link and password to the online survey to the list of contacts by e-mail. The survey was sent in December 2014, with reminders in January 2015 and February 2015. In total, 118 of the 334 invited organization representatives logged in to answer the survey. Out of these, we received 74 completed questionnaires (9 only read the initial instructions, 18 dropped at the first page of the questionnaire, and 17 dropped the survey later without completing the questionnaire). The median time to answer the survey completely was 29 minutes.

---
[2] EFS survey tool: www.unipark.com/en
[3] Softex: http://www.softex.br

### 3.3 Characterization

To provide a summary of the characterization of the responding organizations in Austria and Brazil, we will present information on their company size, used process models, and RE standards. We will also present the roles of the participants within the organizations and their experience in this role. While the data from Austria is more representative to the European context and relies on carefully selected experts in requirements and software quality engineering, we believe that the large data set from Brazil serves as an interesting complement to enable further understanding the investigated phenomena. Concerning size, in Table 1 presents the data from Austria and Table 2 presents the data from Brazil. It is possible to observe that, while in both countries we have small and large organizations, in the Austrian set the medium-sized organizations also play a relevant role representing 33% (cf. 251-500 employees) of the valid answers.

**Table 1.** Size of the organizations surveyed in Austria.

| Size* | No. of Answers | Share [%] |
|---|---|---|
| 1-10 Employees | 2 | 16.68 % |
| 11-50 Employees | 1 | 8.33 % |
| 51-250 Employees | 1 | 8.33 % |
| 251-500 Employees | 4 | 33.33 % |
| 501-1000 Employees | 0 | 0.00 % |
| 1001-2000 Employees | 1 | 8.33 % |
| More than 2000 Employees | 3 | 25.00 % |
| Invalid (missing) answers | 2 | n/a |
| Valid Responses: | 12 | 100.00 % |

* Size including software and other areas.

---

[4] MPS-SW: http://www.softex.br/mpsbr

**Table 2.** Size of the organizations surveyed in Brazil.

| Size* | No. of Answers | Share [%] |
|---|---|---|
| 1-10 Employees | 11 | 15.49% |
| 11-50 Employees | 15 | 21.13% |
| 51-250 Employees | 17 | 23.94 % |
| 251-500 Employees | 5 | 7.04 % |
| 501-1000 Employees | 3 | 4.23 % |
| 1001-2000 Employees | 5 | 7.04 % |
| More than 2000 Employees | 15 | 21.13 % |
| Invalid (missing) answers | 3 | n/a |
| Valid Responses: | 71 | 100.00 % |

* Size including software and other areas.

Regarding the process model, Tables 3 and 4 shows that most of the surveyed organization adopt agile (mainly Scrum-based) process models, followed by iterative and incremental process models and the traditional waterfall model. Note that the respondents could nominate more than one process model typically applied in their organization. A slight difference is that apparently the V-Model XT is more popular in Austria (mentioned by 20.00% of the organizations) than in Brazil (mentioned by 5.41% of the organizations). It is noteworthy that some organizations reported to use more than one process model to handle different types of projects. One explanation for changing process models is that organizations might have to follow a waterfall like model during a bidding procedure while adopting Scrum after formal project assignment.

**Table 3.** Process models used in Austria.

| Process Model | No. of Answers | Share [%] |
|---|---|---|
| Scrum | 6 | 40.00 % |
| Waterfall | 4 | 26.67 % |
| V-Model XT | 3 | 20.00 % |
| Rational Unified Process (RUP) | 1 | 6.67 % |
| Extreme Programming (XP) | 0 | 0.00 % |
| Others* | 4 | 26.67 % |
| Organizations (multiple answers possible): | 14 | 100.00% |

* Others includes project or customer dependent process (2), and other process models based on agile (1) or plan-driven methods (1).

**Table 4.** Process models used in Brazil.

| Process Model | No. of Answers | Share [%] |
|---|---|---|
| Scrum | 45 | 60.81 % |
| Waterfall | 22 | 29.73 % |
| Rational Unified Process (RUP) | 19 | 25.68 % |
| Extreme Programming (XP) | 7 | 9.46 % |
| V-Model XT | 4 | 5.41 % |
| Others* | 11 | 14.86 % |
| Organizations (multiple answers possible): | 74 | 100.00 % |

* Others includes self-adapted process models (4), other iterative and incremental development process models (4) and other process models based on agile methods (3).

Tables 5 and 6 presents the application of RE standards reported by the Austrian and Brazilian respondents. We can observe that in Austria most organizations adopt self-defined standards and few of them base their standards on external regulations and/or software reference models. In Brazil, on the other hand, most of the surveyed organizations follow regulation/reference-model-based standards. This, of course, may have been influenced by the strategy of also distributing the survey to the organizations with valid MPS-SW assessments. Nevertheless, many organizations answered that they follow the standards of the adopted development process and their own standards.

**Table 5.** RE Standards used in Austria.

| RE Standard | No. of Answers | Share [%] |
|---|---|---|
| Self-defined (including artefacts and templates) | 7 | 50.00 % |
| Self-defined (including a process with roles and responsibilities) | 6 | 42.86 % |
| Adopted development process (e.g., RUP, Scrum) | 4 | 28.57 % |
| Self-defined (including a process with deliverables, milestones and phases) | 4 | 28.57 % |
| Regulation (e.g., ITIL) / SW ref. model (e.g., CMMI-Dev) | 2 | 14.29 % |
| None | 0 | 0.00 % |
| Others* | 1 | 7.14 % |
| Organizations (multiple answers possible): | 14 | 100.00 % |

* Others includes project or customer dependent standards (1)

**Table 6.** RE Standards used in Brazil.

| RE Standard | No. of Answers | Share [%] |
| --- | --- | --- |
| Regulation (e.g. ITIL) / SW ref. model (e.g., CMMI-Dev, MPS-SW) | 39 | 52.70 % |
| Adopted development process (e.g., RUP, Scrum) | 25 | 33.78 % |
| Self-defined (including a process with deliverables, milestones and phases) | 19 | 25.68 % |
| Self-defined (including a process with roles and responsibilities) | 18 | 24.32 % |
| Self-defined (including artefacts and templates) | 18 | 24.32 % |
| None | 1 | 1.35 % |
| Organizations (multiple answers possible): | 74 | 100.00 % |

To characterize the participants, the NaPiRE survey collects their roles in the organization and their experience. The roles in Austria and Brazil are shown in Tables 7 and 8. It can be seen that participants in both countries are mainly project managers and business analysts. The main difference is that in Austria the answers are more evenly distributed between the roles, while in Brazil about half of the answers were provided by project managers.

**Table 7.** Roles of the participants in Austria.

| Role | No. of Answers | Share [%] |
| --- | --- | --- |
| Business Analyst | 3 | 25.00 % |
| Project Manager | 2 | 16.67 % |
| Requirements Engineer | 2 | 16.67 % |
| Test Manager / Tester | 2 | 16.67 % |
| Architect | 1 | 8.32 % |
| Others* | 2 | 16.67 % |
| Invalid (missing) answers | 2 | n/a |
| Valid responses: | 12 | 100.00 % |

* Others include trainer and test manager (1), and quality assurance (1).

**Table 8.** Roles of the participants in Brazil.

| Role | No. of Answers | Share [%] |
|---|---|---|
| Project Manager | 32 | 45.07 % |
| Business Analyst | 8 | 11.27 % |
| Developer | 4 | 5.63 % |
| Software Architect | 4 | 5.63 % |
| Test Manager / Tester | 3 | 4.23 % |
| Requirements Engineer | 2 | 2.82 % |
| Others* | 18 | 25.35 % |
| Invalid (missing) | 3 | n/a |
| Valid responses: | 71 | 100.00 % |

* Other informed values include development directors, program managers and portfolio managers (7), quality assurance analysts (7), and people from the software engineering process group (4).

Finally, Tables 9 and 10 show that participants of both countries are highly experienced in their roles, with the majority having more than 3 years of experience.

**Table 9.** Experience of the participants in Austria.

| Experience | No. of Answers | Share [%] |
|---|---|---|
| Expert (more than 3 years) | 9 | 81.82 % |
| Experienced (1 to 3 years) | 2 | 18.18 % |
| Novice (up to 1 year) | 0 | 0.00 % |
| Invalid (missing) | 3 | n/a |
| Valid responses: | 11 | 100.00 % |

**Table 10.** Experience of the participants in Brazil.

| Experience | No. of Answers | Share [%] |
|---|---|---|
| Expert (more than 3 years) | 52 | 73.24 % |
| Experienced (1 to 3 years) | 15 | 21.13 % |
| Novice (up to 1 year) | 4 | 5.63 % |
| Invalid (missing) | 3 | n/a |
| Valid responses: | 71 | 100.00 % |

## 4   Criticality of RE Problems in Austria and Brazil

During the NaPiRE survey, based on a set of 21 precompiled general RE problems listed in the NaPiRE questionnaire [5], participants were asked – according to their expertise – to rank the five most critical requirement issues. The outcomes in Austria and Brazil are shown in Tables 11 and 12. In these tables, we present all issues that were cited among the five most critical requirements issues by at least 20% of the participants. We also show how often each problem was cited and how often it was ranked as the most critical. For instance, Table 11 shows that problem *incomplete/hidden requirements* was cited as one of the five most critical by 9 of the 14 Austrian participants (64.28%) and this issues has been listed as the most critical one by five of them (35.71%).

It is possible to observe that in both countries the most critical reported RE problems are related *incomplete/hidden requirements*, *underspecified requirements*, *communication flaws between the project team and the customer*, and *communication flaws within the project team*. Besides these problems, both tables also share the *moving targets* and *time boxing* problems. We believe that this very similar reported problem profile might be due to using similar process models (mainly Scrum-based, cf. Table 3 and Table 4). Differences in the criticality were observed in the "*stakeholders with difficulties in separating requirements from previously known solution designs problem*", which was cited by more than 20% of the participants in Austria, but not in Brazil. On the other hand, the problems "*insufficient support by customer*" and "*inconsistent requirements*" were cited by more than 20% of the participants in Brazil, but not in Austria.

**Table 11.** Most critical RE problems in Austria.

| # | RE Problems and Issues | Cited* | | Ranked #1* | |
|---|---|---|---|---|---|
| | | No. | % | No. | % |
| **1** | **Incomplete and/or hidden requirements** | **9** | **64.28 %** | **5** | 35.71% |
| 2 | Underspecified requirements that are too abstract and allow for various interpretations | 4 | 26.67% | 1 | 7.14% |
| 3 | Communication flaws within the project team | 4 | 26.67% | 1 | 7.14% |
| 4 | Communication flaws between the project team and the customer | 3 | 21.42% | 1 | 7.14% |
| 4 | Moving targets (changing goals, business processes and/or req.) | 3 | 21.42% | 1 | 7.14% |
| 4 | Stakeholders with difficulties in separating reqs from previously known solution designs | 3 | 21.42% | 3 | 21.43% |
| 4 | Time boxing / Not enough time in general | 3 | 21.42% | 1 | 7.14% |

\* The probabilities were calculated based on the overall amount of 14 participants.

**Table 12.** Most critical RE problems in Brazil.

| # | RE Problems and Issues | Cited* No | Cited* % | Ranked #1* No | Ranked #1* % |
|---|---|---|---|---|---|
| 1 | Communication flaws between the project team and the customer | 32 | 43.24% | 9 | 12.16% |
| **2** | **Incomplete and/or hidden requirements** | **31** | **41.89%** | **12** | **16.22%** |
| 2 | Underspecified requirements that are too abstract and allow for various interpretations | 31 | 41.89% | 3 | 4.05% |
| 4 | Communication flaws within the project team | 26 | 35.14% | 5 | 6.67% |
| 5 | Insufficient support by customer | 21 | 28.38% | 5 | 6.76% |
| 6 | Inconsistent requirements | 18 | 24.32% | 2 | 2.70% |
| 7 | Time boxing / Not enough time in general | 17 | 22.97% | 1 | 1.35% |
| 8 | Moving targets (changing goals, business processes and/or req.) | 15 | 20.27% | 5 | 6.67% |

* The probabilities were calculated based on the overall amount of 74 participants.

In this paper, we focus on the specific problem of *incomplete/hidden requirements*. According to Tables 11 and 12 this issue is highly relevant for both contexts, being the most cited problem in Austria and the second most cited problem in Brazil. Moreover, the majority of respondents have cited this issue as the most critical one in both countries (see the last columns of Tables 11 and 12).

## 5 Analyzing the Incomplete/Hidden Requirements Problem

Considering the specific problem of incomplete/hidden requirements, Tables 13 and 14 show how survey respondents from Austria and Brazil judge its applicability to their own projects (participants were asked to judge the applicability of all the precompiled RE problems). In this question, *incomplete* and *hidden* requirements were analyzed separately, which would not make sense for the question to rank the most critical ones discussed in the previous section, as these problems are often similar (requirements are often incomplete because there are hidden requirements which were not specified) and should therefore not be counted twice in a ranking.

It can be observed that in both countries most of the respondents consider the problem applicable/relevant to their own projects, with more than 75% and 65% agreeing or partially agreeing on its relevance in Austria and in Brazil, respectively. In fact, the judgements for both items, *incomplete* and *hidden*, were almost similar in each of the countries, which reinforces the decision of analyzing them together as *incomplete/hidden requirements* when discussing the most relevant problems and their causes.

**Table 13.** Applicability/relevance of incomplete and hidden requirements to projects of Austrian respondents.

| Problem | Disagree | Partially Disagree | Neutral | Partially Agree | Agree | Valid responses |
|---|---|---|---|---|---|---|
| Incomplete | 0 (0.00%) | 1 (7.69%) | 2 (15.38%) | 7 (53.85%) | 3 (23.08%) | 13 (100 %) |
| Hidden requirements | 0 (0.00%) | 0 (0.00%) | 2 (15.38%) | 6 (46.15%) | 5 (38.46%) | 13 (100 %) |

**Table 14.** Applicability/relevance of incomplete and hidden requirements to projects of Brazilian respondents.

| Problem | Disagree | Partially Disagree | Neutral | Partially Agree | Agree | Valid responses |
|---|---|---|---|---|---|---|
| Incomplete | 8 (11.94%) | 2 (2.99%) | 12 (17.91%) | 17 (25.37%) | 28 (41.79%) | 67 (100 % |
| Hidden requirements | 7 (10.45%) | 0 (0.00%) | 14 (20.90%) | 18 (26.87%) | 28 (41.79%) | 67 (100 % |

After selecting the five most critical RE problems, respondents were asked to provide what they believe of being the main causes for each of the problems. They provided the causes in an open question format, with one open question for each of the previously selected RE problems.

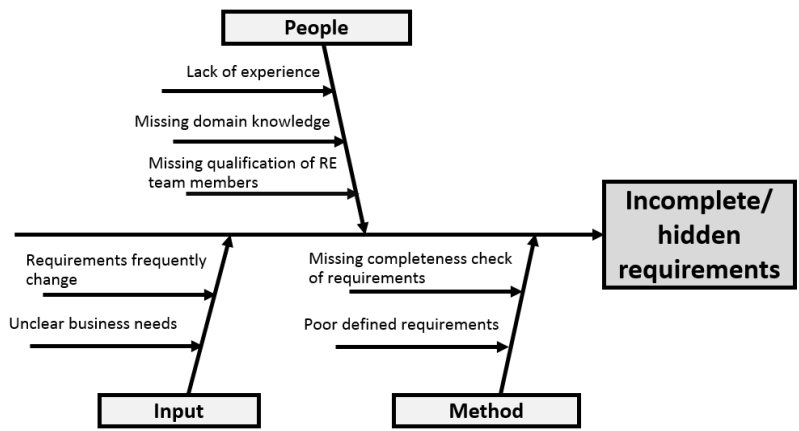

**Fig. 1.** Austrian cause-effect diagram for incomplete/hidden requirements.

Six of the nine respondents from Austria that reported *incomplete/hidden requirements* among the most critical ones also listed causes for this problem. We analyzed their textual cause descriptions, using the coding terms used for the German

NaPiRE trial as a starting point and decided to add new terms only when strictly needed. As a result, we identified 7 causes (each one cited once) and no new coding terms were needed. Then, we represented these causes in a cause-effect diagram [16], using the categories suggested in [6]: *input*, *method*, *organization*, *people*, and *tools*. The resulting cause-effect diagram is shown in Figure 1.

We repeated the same process for the Brazilian data, in which 27 out of the 31 that reported incomplete/hidden requirements among the most critical ones also listed causes for this problem. We identified 18 different causes in the textual descriptions (in this case, the coding terms were slightly extended – adding four new terms – due to textual descriptions that could not be mapped to the previously provided terms). Given the size of this data set, we also counted the frequency in which each cause was cited (at all we had 35 cause citations).

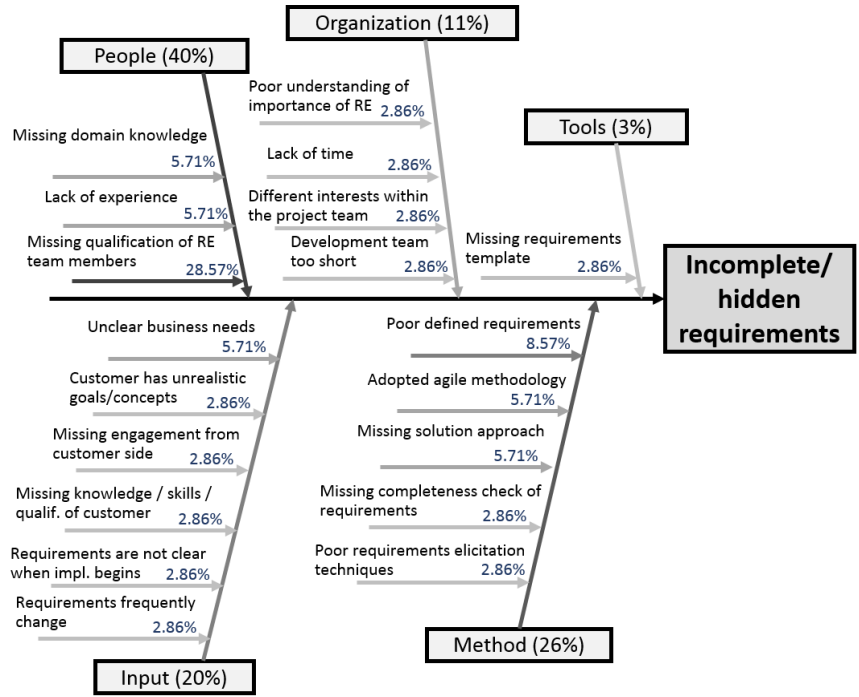

**Fig. 2.** Brazilian probabilistic cause-effect diagram for incomplete/hidden requirements.

With this additional information on the frequency, we were able to build a probabilistic cause-effect diagram [17][18], which enables identifying the most common causes based on probabilistic percentages (in this case, their frequencies). Figure 2 extends the traditional cause-effect diagram [16] by (a) showing the probabilities for each possible cause to lead to the analyzed problem, and (b) representing the causes using grey tones, where causes with higher probability are shown closer to the center and in darker tones. The resulting probabilistic cause-effect diagram is shown in Figure 2. We believe that this representation complements the information on causes reported for the problem in Austria. In fact, the causes reported in Austria are contained in the causes reported in Brazil, with additional causes and information on their frequency based on a larger sample. In fact, most of the most frequently cited causes in Brazil shown in Figure 2 were also identified in the results of the Austrian survey. According to the survey responses we highlight the *missing qualification of RE team members*, *lack of experience*, *missing domain knowledge*, *unclear business needs* and *poor defined requirements* as the main causes.

To address the first three of these causes, related to the *people* category, aiming prevention, we recommend training on best RE practices, selecting highly experienced requirements analysts and involving domain experts and/or providing appropriate training on the application domain. For cases were the lack of domain knowledge plays a significant role, we also recommend some specific domain immersive elicitation techniques, such as ethnography. *Unclear business needs* can be addressed by applying business case analysis that helps fostering discussions and clarifying business objectives and values and by facilitating a stronger involvement and clear communications of the customer. In context of RE joint RE workshops in collaboration with the customer might help to precisely identify the real business needs. Finally, the *poor defined requirements* could be addressed by providing a detailed requirements specification template and conducting peer reviews with appropriate inspection methods (e.g., checklists or reading techniques), ideally involving different stakeholders (e.g., users, designers, and testers) in the verification and validation process. These counter measures represent a set of initial strategies based on the experience of the study team, i.e., the authors.

However, during the NaPiRE survey, candidate measures to address these issues have been collected from survey participants. Table 15 and 16 presents an overview on risk and RE issue mitigation actions, reported by the participants in Austria and Brazil. These mitigation action can serve as an additional input (from industry projects) to investigate best practices to prevent the incomplete/hidden requirements problem. However, more detailed analysis is required to investigate (a) which mitigation actions are most promising to improve the incomplete/hidden requirements problem and (b) how to support engineers in better addressing these issue.s

**Table 15.** Mitigation actions for incomplete/hidden requirements reported in Austria.

| **Mitigation Actions for Incomplete/Hidden Requirements** |
| --- |
| Having testers testing requirements. |
| Increased efforts during the review process. |
| After project retrospective with project team. |
| Checklists for requirements. |

**Table 16.** Mitigation actions for incomplete/hidden requirements reported in Brazil.

| **Mitigation Actions for Incomplete/Hidden Requirements** |
| --- |
| Improve the documentation and conduct more meetings with the developers to detect analysis defects. |
| Hire or specialize a requirements analyst. |
| Creating templates. |
| Creation of a DoR (Definition of Readiness) for the team. |
| Invest more time in requirements specification, using scenarios and prototypes to gather requirements more completely. |
| Peer reviews involving testes. |
| Invest more effort in requirements validation using prototypes. |
| Peer reviews involving developers. |
| Provide training to the RE team. |
| Process models. |
| Avoiding including incomplete requirements, when already known to be incomplete, in development sprints. |
| Prototyping; technical reviews and consensus meetings. |
| Improve the analysis to be more detailed. |
| More frequent meetings with the customer to align expectations. |
| Requirements reviews and frequent releases. |
| Improving the quality of the requirements documentation, or improving elicitation methods. |
| Developing requirements according to suggestions of the MPS-SW reference model. |
| Improvement of the artefacts; adoption of software inspections. |
| Standardizing the requirements specifications, using a validation checklist and peer reviews. |
| Training, mentoring, selecting professionals with an adequate profile, a highly skilled team. |
| Provide training to the RE team. |
| Reviewing the RE processes. |
| The customer should have a better understanding of the problem; requirements verification with all stakeholders (applying Perspective-Based Reading). |

## 6 Concluding Remarks

Many projects fail due to problems in RE. In this paper, we further analyzed a specific and relevant RE problem: *incomplete/hidden requirements*. Therefore, we used the data of the NaPiRE survey replications we conducted in Austria and Brazil. We provided the basic characterization of the responding organizations (14 in Austria and 74 in Brazil), which include small, medium and large sized companies, conducting both, plan-driven and agile development. Thereafter, we characterized the criticality of the selected RE problem. Results showed that in both countries the survey respondents considered it one of the most critical RE problems (#1 in Austria and #2 in Brazil) and reported that it is applicable and relevant to their projects.

To provide further knowledge on the causes of this problem, we compiled all the causes reported in Austria into a cause-effect diagram and the causes reported in the large Brazilian sample into a probabilistic cause-effect diagram. Most commonly reported causes were *missing qualification of RE team members*, *lack of experience*, *missing domain knowledge*, *unclear business needs* and *poor defined requirements*.

Based on these causes, we discussed solution options on how to address them in order to prevent incomplete/hidden requirements in future projects. Furthermore, we compiled the lists of mitigation actions cited by the survey respondents from Austria and Brazil, which may serve as additional input for preventing the problem.

We believe that, from a practical point of view, this paper provides further insights into common causes of incomplete/hidden requirements and on how to prevent this problem.

**Future work** includes a more detailed analysis of NaPiRE Austria and NaPiRE Brazil surveys with regard to other RE problems, and to triangulate our results with data from other countries where NaPiRE was performed to increase the validity and reliability of the results achieved.

**Acknowledgments.** The authors would like to thank the NaPiRE community for their support. Thanks also to the Brazilian research council (CNPq) for financial support (grant #460627/2014-7). Part of this work was also supported by the Christian Doppler Forschungsgesellschaft, the Federal Ministry of Economy, Family and Youth, and the National Foundation for Research, Technology and Development, Austria.